\documentclass[12pt]{iopart}
\usepackage{epsf}
\begin{document}
\title{Constraints on alternative models to dark energy}
\author{Yungui Gong}
\address{Institute of Applied Physics and
College of Electronic Engineering, Chongqing University of Post
and Telecommunication, Chongqing 400065, China}
\ead{gongyg@cqupt.edu.cn}
\author{Chang-Kui Duan}
\address{Institute of Applied Physics and
College of Electronic Engineering, Chongqing University of Post
and Telecommunication, Chongqing 400065, China}
\ead{duanck@cqupt.edu.cn}
\begin{abstract}
The recent observations of type Ia supernova strongly support that
the universe is accelerating now and decelerated in the recent
past. This may be the evidence of the breakdown of the standard
Friedmann equation. We consider a general modified Friedmann
equation. Three different models are analyzed in detail. The
current supernova data and the Wilkinson microwave anisotropy
probe data are used to constrain these models. A detailed analysis
of the transition from  the deceleration phase to the acceleration
phase is also performed.
\end{abstract}
 \maketitle

\parindent=4ex
\section{Introduction}
The recent observations of Type Ia supernova indicate that the
expansion of the universe is speeding up \cite{sp99}.
Observational results also provide the evidence of a decelerated
universe in the recent past \cite{agr}. On the other hand, the
cosmic background microwave (CMB) observations indicate that the
universe is spatially flat as predicted by the inflationary models
\cite{pdb00}. A dark energy component with negative pressure
behavior which dominates the universe, was proposed to explain a
flat and accelerating universe. One simple candidate of dark
energy is the cosmological constant. However, there are some
problems with the cosmological constant although cold dark matter
cosmological constant models are consistent with the current
observations. Why is the cosmological constant so small and not
zero? Why does the cosmological constant become significant now?
The quintessence models avoid some of the problems \cite{cds98}.
There are also other models, like tachyon filed as dark energy
\cite{padmt03}. But the property of dark energy is still
mysterious. One logical possibility is that the standard Friedmann
equation may need to be modified. In this scenario, the universe
is dominated by ordinary pressureless matter, but the law of
gravity and the standard Friedmann equation are modified. The idea
of modifying the law of gravity is not new. The modified Newtonian
Dynamics (MOND) was first used to explain the rotation curve in
place of dark matter \cite{mond,mond1}. In MOND, the Newtonian
gravity $M/r^2$ is replaced with $M/r^2+\sqrt{M}/r$. Since $M/r^2$
gives the standard Freidmann equation $H^2\sim \rho$,
$M/r^2+\sqrt{M}/r$ may provide a modified Friedmann equation
$H^2\sim \rho^{2/3}\ln \rho+ \rho^{2/3}$ \cite{mond1}. Recall that
the brane cosmology gives a non-standard Friedmann equation
$H^2\sim \rho +\rho^2$ \cite{bdl,rs99,gong00}. Along this line of
reasoning, Freese and Lewis recently proposed the Cardassian
expansion in which the universe is dominated by the ordinary
matter and the Friedmann equation becomes $H^2\sim \rho+\rho^n$
\cite{freese02}. The Cardassian model was later generalized to a
more general form $H^2\sim g(\rho)$ \cite{chung}. In addition,
several authors modified the Friedmann equation as $H^2+H^\alpha
\sim \rho$ motivated by theories with extra dimensions
\cite{dvali00,lue03a}. In this paper, we first consider a model
which is equivalent to the generalized Chaplygin gas model
\cite{kamenshchik} in terms of dynamical evolution. Then we
consider the generalized Cardassian model. At last we consider a
model proposed by Dvali, Gabadadze and Porrati (DGP)
\cite{dvali00}. We use the ten new supernovae at $z=0.36-0.86$
\cite{raknop03} and the the first year Wilkinson Microwave
Anisotropy Probe (WMAP) temperature (TT) and temperature
polarization cross correlation (TE) data \cite{ghinshaw} to
constrain these models. We also investigate the transition from
the decelerated phase to the accelerated phase.

For a spatially flat, isotropic and homogeneous universe with both
an ordinary pressureless dust matter and a minimally coupled
scalar field $Q$ sources, the Friedmann equations are
\begin{eqnarray}
\label{cos1} H^2=\left({\dot{a}\over a}\right)^2={8\pi
G\over 3}(\rho_{\rm m}+\rho_{\rm Q}),\\
\label{cos2}
{\ddot{a}\over a}=-{4\pi G\over
3}(\rho_{\rm m}+\rho_{\rm Q}+3p_{\rm Q}),\\
\label{cos3} \dot{\rho_{\rm Q}}+3H(\rho_{\rm Q}+p_{\rm Q})=0,
\end{eqnarray}
where dot means derivative with respect to time, $\rho_{\rm
m}=\rho_{\rm m0}(a_0/a)^3$ is the matter energy density, a
subscript 0 means the value of the variable at present time,
$\rho_{\rm Q}=\dot{Q}^2/2+V(Q)$, $p_{\rm Q}=\dot{Q}^2/2-V(Q)$ and
$V(Q)$ is the potential of the quintessence field. The modified
Friedmann equations (MFE) for a spatially flat universe are
\begin{eqnarray}
\label{cosa} H^2=H_0^2g(x),\\
\label{cosb} {\ddot{a}\over a}=H^2_0g(x)
-{3H^2_0x\over 2}g'(x)\left({\rho+p\over \rho}\right),\\
\label{cosc} \dot{\rho}+3H(\rho+p)=0,
\end{eqnarray}
where $x=8\pi G\rho/3H^2_0=x_0(1+z)^3$ during the matter dominated
epoch, $1+z=a_0/a$ is the redshift parameter, $g(x)=x+\cdots$ is a
general function of $x$ and $g'(x)=dg(x)/dx$. From
phenomenological point of view, four dimensional gravity is
modified so that we get a general function $g(x)$. Chung and
Freese argued that a general $g(x)$ is possible if our universe as
a three brane is embedded in five dimensional spacetime
\cite{chung}. For example, $g(x)\sim x+x^2$ in Brane cosmology,
$g(x)\sim x+x^n$ in Cardassian model. On the other hand, we can
think the additional terms $g(x)-x$ as dark energy component. For
the Cardassian model, the additional term $x^n$ can be mapped to a
dark energy component with constant equation of state parameter
$\omega_{\rm Q}=n-1$. In general, we get the following
relationship between the dark energy equation of state parameter
and $g(x)$
\begin{equation}
\label{darkom} \omega_{\rm Q}={xg'(x)-g(x)\over g(x)-x}.
\end{equation}
The equivalent dark energy potential can be found from the
following equations
\begin{eqnarray}
\dot{Q}^2=\rho_{\rm m}[g'(x)-1],\\
V(Q)={3H^2_0\over 8\pi G}[ g(x)-0.5x-0.5xg'(x)].
\end{eqnarray}
For instance, if $g(x)\sim x +x^n$, we find that $V(Q)\sim
[\sinh(AQ+B)]^{2n/(n-1)}$. Note that the universe did not start to
accelerate when the other terms in $g(x)$ started to dominate. The
linear density perturbation of this model is given by
\begin{equation}
\label{perturb} \ddot{\delta}+2\bar{H}\dot{\delta}=4\pi
G\bar{\rho}\delta[g'(\bar{x})+3\bar{x}g''(\bar{x})].
\end{equation}
For the matter dominated flat universe, $\rho=\rho_{\rm m}$ and
$p=p_{\rm m}=0$. Let $\Omega_{\rm m0}=8\pi G\rho_0/3H^2_0$, then
$x_0=\Omega_{\rm m0}$, $g(x_0)=1$. In general, $x=\Omega_{\rm
m0}(1+z)^3+\Omega_{\rm r0}(1+z)^4$, where $\Omega_{\rm
r0}=8.35\times 10^{-5}$ is the current radiation component
\cite{DNSpergel}.

\section{Analytical Method}
The location of the $m$-th peak of the CMB power spectrum is
parameterized as \cite{doran}
\begin{equation}
\label{peak1} l_m=(m-\phi_ m)l_{\rm A}, \end{equation} where the
acoustic scale $l_A$ is
\begin{equation}
l_{\rm A}={\pi\over \bar{c_s}}{\tau_0-\tau_{\rm ls}\over \tau_{\rm
ls}}={\pi\over \bar{c_s}}{\int^{z_{\rm ls}}_0 (g[\Omega_{\rm
m0}(1+z)^3+\Omega_{\rm r0}(1+z)^4])^{-1/2} dz\over
\int^{\infty}_{z_{\rm ls}} (g[\Omega_{\rm m0}(1+z)^3+\Omega_{\rm
r0}(1+z)^4])^{-1/2} dz},
\end{equation}
$\bar{c_s}=0.52$, the conformal time at last scattering $\tau_{\rm
ls}$ and today $\tau_0$ are
\begin{eqnarray}
\tau_{\rm ls}=\int^{\tau_{\rm ls}}_0 d\tau=\int^{\infty}_{z_{\rm
ls}} {dz\over
a_0 H_0\sqrt{g[\Omega_{\rm m0}(1+z)^3+\Omega_{\rm r0}(1+z)^4]}},\\
\label{peak2} \tau_0=\int^{\tau_0}_0 d\tau=\int^{\infty}_0
{dz\over a_0 H_0\sqrt{g[\Omega_{\rm m0}(1+z)^3+\Omega_{\rm
r0}(1+z)^4]}}.
\end{eqnarray}
The recent WMAP results give the positions of the first two
acoustic peaks as $l_{p1}=220.1\pm 0.8$ and $l_{p2}=546\pm 10$,
respectively \cite{ghinshaw}. The third peak is given by the
BOOMERanG measurements as $l_{p3}=845^{+12}_{-25}$
\cite{bernardis}. We first use equations
(\ref{peak1})-(\ref{peak2}) with $\phi_1=0.3074$, $\phi_2=0.2819$
and $\phi_3=0.341$ to constrain the models considered below, then
we use the full 1350 WMAP TT and TE data \cite{ghinshaw} by a
modified CMBFAST code version 4.5.1 \cite{cmbfast} to constrain
the parameters. For the fit to full WMAP data, a scalar power
spectrum with normalization 0.833, spectral index 0.93 and running
index slope -0.031 are assumed. Other cosmological parameters are
chosen as follows: $h=0.71$, $\Omega_{\rm b}=0.044$, $T_{\rm
cmb}=2.725$, Helium abundance $Y_{\rm He}= 0.24$, Number of
massless neutrinos is 3.04 and $g^*=10.75$.

The luminosity distance $d_{\rm L}$ is defined as
\begin{equation}
\label{lumin} d_{\rm L}(z)=a_0(1+z)\int^{t_0}_t {dt'\over
a(t')}={1+z\over H_0}\int^z_0 {du\over \sqrt{g[\Omega_{\rm
m0}(1+u)^3]}}.
\end{equation}
The apparent magnitude redshift relation becomes
\begin{eqnarray}
\label{magn}
 m(z)&=&M+5\log_{10}d_{\rm L}(z)
+25=\mathcal{M}+5\log_{10}\mathcal{D}_{\rm L}(z) \nonumber \\
&=& \mathcal{M}+5\log_{10}\left[(1+z)\int^z_0 {du\over
\sqrt{g[\Omega_{\rm m0}(1+u)^3]}}\right],
\end{eqnarray}
where $\mathcal{D}_{\rm L}(z)=H_0d_{\rm L}(z)$, $M$ is the
absolute magnitude and $\mathcal{M}=M-5\log_{10}H_0+25$. The
nuisance parameter $\mathcal{M}$ can be determined from the low
redshift limit at where $\mathcal{D}_{\rm L}(z)=z$. The parameters
in the models are determined by minimizing
\begin{equation}
\label{lrmin} \chi^2=\sum_i{[m_{\rm obs}(z_i)-m(z_i)]^2\over
\sigma^2_i},
\end{equation}
where $\sigma_i$ is the total uncertainty in the observation. We
use equations (\ref{lumin}) and (\ref{magn}) to fit the ten new
supernova data with host galaxy correction \cite{raknop03}. The
nuisance parameter $\mathcal{M}$ is marginalized when we fit the
supernova data.

The transition from deceleration to acceleration happens when the
deceleration parameter $q=-\ddot{a}/aH^2=0$. From equations
(\ref{cosa}) and (\ref{cosb}), we have
\begin{eqnarray}
\label{trans} g[\Omega_{\rm m0}(1+z_{q=0})^3]={3\over
2}\Omega_{\rm m0}(1+z_{q=0})^3g'[\Omega_{\rm m0}(1+z_{q=0})^3],\\
\label{qparam} q_0={3\over 2}\Omega_{\rm m0}g'(\Omega_{\rm m0})-1.
\end{eqnarray}

\section{Chaplygin Gas Model}
\begin{figure}[htb]
\begin{center}
\epsfxsize=5in \epsffile{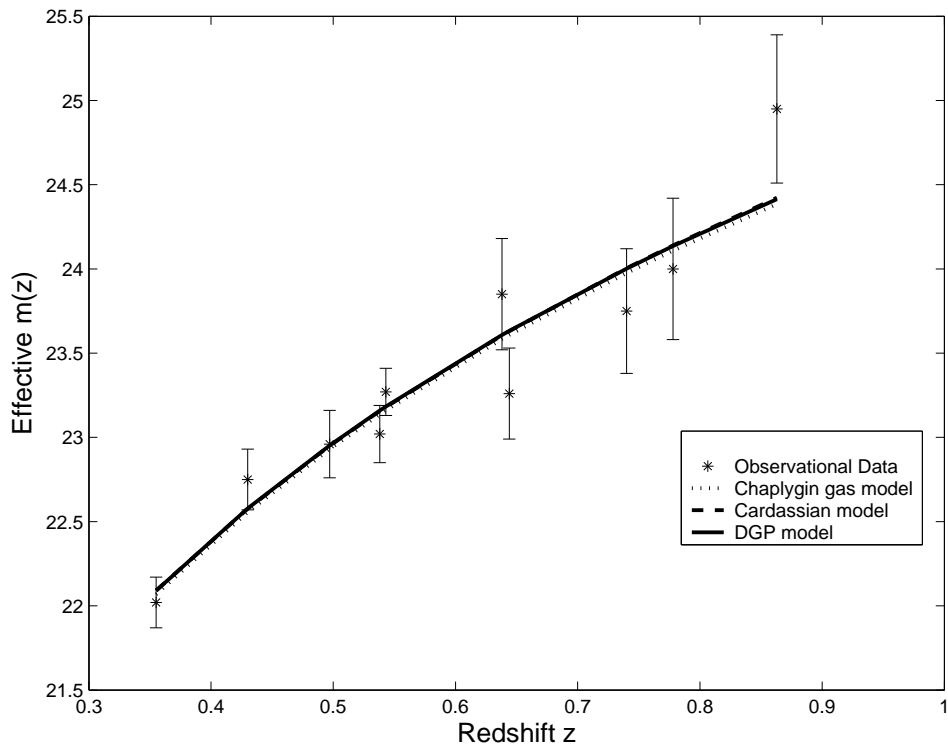}
\end{center}
 \caption{The Best fit curve to the supernova data.
Generalized Chaplygin gas: $\Omega_{\rm m0}=0.27$, $A_s=0.94$ and
$\beta=1.0005$; Generalized Cardassian model: $\Omega_{\rm m0}\sim
0$, $\alpha=0.23$ and $n=0.49$; DGP model: $\Omega_{\rm m0}=0.21$}
\label{bestfit}
\end{figure}

In the framework of MFE, the generalized Chaplygin gas model $p_c=-A/\rho_c^\alpha$ becomes
$$g(x)=x+\Omega_{\rm Q0}[A_s+(1-A_s)(x/\Omega_{\rm m0})^\beta]^{1/\beta},$$
where $\Omega_{\rm Q0}=1-\Omega_{\rm m0}-\Omega_{\rm r0}$,
$\beta=1+\alpha$ and $A_s=(8\pi G/3H^2_0\Omega_{\rm Q0})^\beta A$.
To recover the standard Friedmann equation at early times, we need
$A_s\sim 1$. When $A_s=1$, the model becomes a standard
$\Lambda$-model. Now we have
\begin{eqnarray}
\label{confta}
\fl \int^z_0 {du\over \sqrt{g[\Omega_{\rm m0}(1+u)^3+\Omega_{\rm r0}(1+u)^4]}}  \nonumber\\
\fl =\int^z_0{du\over \sqrt{f(u)+\Omega_{\rm Q0}\{A_s
+(1-A_s)(1+u)^3+\Omega_{\rm r0} (1+u)^4/\Omega_{\rm
m0}]^\beta\}^{1/\beta}}},
\end{eqnarray}
where $f(u)=\Omega_{\rm m0}(1+u)^3+\Omega_{\rm r0}(1+u)^4$. Since
$g'(x)=1+\Omega_{\rm Q0}(1-A_s)[A_s+(1-A_s)(x/\Omega_{\rm
m0})^\beta]^{1/\beta-1}(x/\Omega_{\rm m0})^\beta$, together with
equations (\ref{trans}) and (\ref{qparam}), we have
\begin{eqnarray}
\fl {\Omega_{\rm m0}\over 2\Omega_{\rm
Q0}}(1+z_{q=0})^3[A_s+(1-A_s)(1+z_{q=0})^{3\beta}]^{1-1/\beta}
=A_s-{1\over 2}(1-A_s)(1+z_{q=0})^{3\beta},\\
q_0={1\over 2}-{3\over 2}A_s(1-\Omega_{\rm m0}),
\end{eqnarray}
The first three peaks in CMB power spectrum favor a cosmological
constant model with $A_s=1$ or $\beta=1$. The best fit to the WMAP
TT and TE data is $\Omega_{\rm m0}=0.26$, $A_s=0.999$ and
$\beta=1.43$ with $\chi^2=1448.3$. The ten supernova data also
favor a cosmological constant model with $A_s=1$ or $\beta=1$. By
using the best fit parameters to WMAP data, we get $z_{q=0}=0.78$
and $q_0=-0.61$. Because $A_s=0.999$, so $g(x)\approx x$ at early
times. The best fit result to the supernova data is shown in
figure \ref{bestfit} and the WMAP TT power spectrum with the best
fit parameters is plotted in figure \ref{wmapfit}.

\section{Cardassian Model}
We take the generalized Cardassian model
$$g(x)=x[1+Bx^{\alpha(n-1)}]^{1/\alpha},$$
where $B=(\Omega_{\rm m0}^{-\alpha}-1)/\Omega_{\rm
m0}^{\alpha(n-1)}$, $\alpha>0$ and $n<1-1/3(1-\Omega^\alpha_{\rm
m0})$. At early times, $g(x)\sim x$, so the standard cosmology is
recovered. When $n=0$,
$g(x)=B^{1/\alpha}(1+x^\alpha/B)^{1/\alpha}$. For the special case
$\alpha=1$ and $n=0$, $g(x)=x+B$ which is the standard cosmology
with a cosmological constant. If we take $\alpha=1$ and $n=1/2$,
then we have $g(x)=x+B\sqrt{x}$. If we think the generalized
Cardassian model as ordinary Freidmann universe composed of matter
and dark energy, we can identify the following relationship for
the parameters in the Cardassian and quintessence models
$$\omega_{\rm Q0}={(n-1)(1-\Omega_{\rm m0}^\alpha)\over 1-\Omega_{\rm m0}}.$$
The generalized Cardassian model gives
\begin{eqnarray}
\label{modb}
g'(x)=[1+Bx^{\alpha(n-1)}]^{1/\alpha}+(n-1)Bx^{\alpha(n-1)}[1+Bx^{\alpha(n-1)}]^{1/\alpha-1},\\
\label{conftb}
\fl \int^z_0 {du\over \sqrt{g[\Omega_{\rm m0}(1+u)^3+\Omega_{\rm r0}(1+u)^4]}} \nonumber\\
\fl =\int^z_0 {du\over \{f(u)[1+(\Omega_{\rm
m0}^{-\alpha}-1)[(1+u)^3+\Omega_{\rm r0}
 (1+u)^4/\Omega_{\rm m0}]^{\alpha(n-1)}]^{1/\alpha}\}^{1/2}}.
\end{eqnarray}
Combining equation (\ref{modb}) with equations (\ref{trans}) and
(\ref{qparam}), we get
\begin{eqnarray}
1+z_{q=0}=[(\Omega_{\rm m0}^{-\alpha}-1)(2-3n)]^{1/3\alpha(1-n)},\\
q_0={1\over 2}+{3\over 2}(n-1)(1-\Omega_{\rm m0}^\alpha).
\end{eqnarray}
The best fit parameters to the first three peaks in CMB power
spectrum are: $\Omega_{\rm m0}=0.28$, $n=-0.032$ and $\alpha=0.5$
for the generalized Cardassian model and $\Omega_{\rm m0}=0.26$
and $n=-0.12$ for the Cardassian model. For the Cardassian model
$\alpha=1$, the best fit prameters to the full WMAP TT and TE data
are $\Omega_{\rm m0}=0.25$ and $n=0.02$ with $\chi^2=1461.0$. The
best fit parameters to the ten supernovae data are: $\Omega_{\rm
m0}\sim 0$, $n=0.51$ and $\chi^2=6.66$. For the generalized
Cardassian model, the best fit parameters to the ten supernovae
data are: $\Omega_{\rm m0}\sim 0$, $n=0.51$, $\alpha=0.54$ and
$\chi^2=6.66$. In fact, the data is not sensitive to $n$ and
$\alpha$. We get almost the same value of $\chi^2$ with different
$\alpha$ and $n$. So the supernova sample is too sparse to
determine the parameters in generalized Cardassian model. By using
$\Omega_{\rm m0}=0.25$ and $n=0.02$, we get $q_0=-0.6$,
$z_{q=0}=0.82$ and $\omega_{\rm Q0}=-0.98$. The best fit curve to
the supernova data is shown in figure \ref{bestfit}. The CMB TT
power spectrum with the best fit Cardassian parameters is drawn in
figure \ref{wmapfit}.

\section{DGP Model}
\begin{figure}[htb]
\begin{center}
\epsfxsize=5in \epsffile{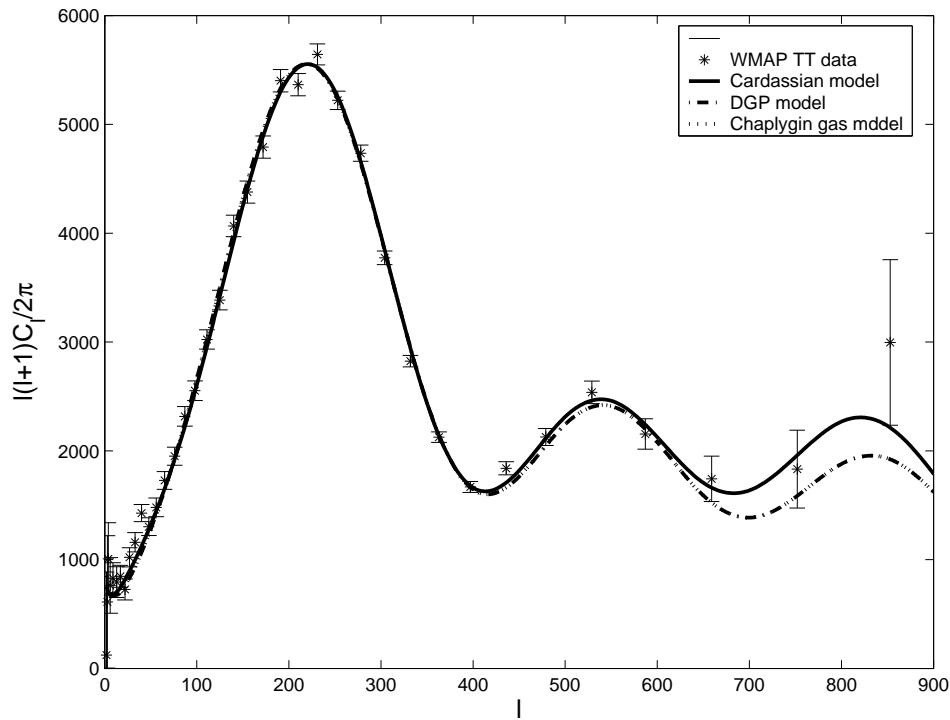}
\end{center}
\caption{The Best fit curve to wmap data. The dashed line is for
generalized Chaplygin gas model with $\Omega_{\rm m0}=0.26$,
$A_s=0.999$ and $\beta=1.43$. The Solid line is for Cardassian
model with $\Omega_{\rm m0}=0.25$ and $n=0.02$; The dash dotted
line is for the DGP model with $\Omega_{\rm m0}=0.174$.}
\label{wmapfit}
\end{figure}

The model is $g(x)=[a+\sqrt{a^2+x}]^2$
\cite{dvali00}, where $a=(1-\Omega_{\rm m0})/2$. At high redshift,
$g(x)\sim x$. So the standard cosmology is also recovered in the
early times. The conformal time is
\begin{equation}
\label{conftc} \fl \int^z_0 {du\over \sqrt{g[\Omega_{\rm
m0}(1+u)^3+\Omega_{\rm r0}(1+u)^4]}}=\int^z_0 {du\over
a+\sqrt{a^2+\Omega_{\rm m0}(1+u)^3+\Omega_{\rm r0}(1+u)^4}}.
\end{equation}
For this model, we find that $q_0$ and the transition redshift
$z_{q=0}$ from decelerated expansion to accelerated expansion are
\begin{eqnarray}
1+z_{q=0}=\left[{2(1-\Omega_{\rm m0})^2\over \Omega_{\rm m0}}\right]^{1/3},\\
q_0={2\Omega_{\rm m0}-1\over 1+\Omega_{\rm m0}}.
\end{eqnarray}
In this model, we have only one free parameter $\Omega_{\rm m0}$.
The locations of the first three peaks in CMB power spectrum give
that $\Omega_{\rm m0}=0.179$. The best fit parameter to the WMAP
TT and TE data is $\Omega_{\rm m0}=0.174$ with $\chi^2=1485.6$.
The best fit parameter to the supernova data is $\Omega_{\rm
m0}=[0,\ 1.0] $ centered at $0.21$ with $\chi^2=6.67$. If we map
the additional term in the right hand side of the Friedmann
equation to dark energy component, we have $\omega_{\rm
Q0}=-1/(1+\Omega_{\rm m0})=-0.83$, $q_0=-0.48$ and $z_{q=0}=0.81$
if we take $\Omega_{\rm m0}=0.21$. The best fit curve to the
supernova data is shown in figure \ref{bestfit} and the best fit
curve to the WMAP data is shown in figure \ref{wmapfit}.

\section{Discussions}

One attractive feature of the generalized Chalygin gas model is
that it can be considered as a unified dark matter and dark energy
model. In the unified scenario, both the dark energy and dark
matter components are modelled by the generalized Chaplygin gas.
In this paper, we are interested in modelling the generalized
Chaplygin gas as dark energy only. We found the best fit
parameters in the generalized Chaplygin gas model are $\Omega_{\rm
m0}=0.26$, $A_s=0.999$ and $\beta=1.43$. Cunha, Alcaniz and Lima
found that $A_s>0.73$ by using the combined x ray data of galaxy
clusters and supernova data and taking a prior $\Omega_{\rm
m0}=0.3$ \cite{kamenshchik}. Amendola \etal found that $1\le\beta<
1.2$ and $0.8< A_s< 1$ by using WMAP TT data \cite{kamenshchik}.
In the context of MFE, the generalized Chaplygin gas model tends
to be the $\Lambda$-model in order to recover the standard
cosmology at early times. Our results show that the generalized
Chaplygin gas model is almost the same as the $\Lambda$-model. By
using the best fit parameter, we get $z_{q=0}=0.78$.

For the Cardassian model, the WMAP TT and TE data give that
$\Omega_{\rm m0}=0.25$ and $n=0.02$. The 10 supernova data gives
that $\Omega_{\rm m0}\sim 0$ and $n=0.51$. The error from
supernova data fit is also large. Zhu and Fujimoto also found low
$\Omega_{\rm m0}$ from supernova data \cite{zhu03}. Sen and Sen
found that $0.31< n < 0.44$ and $0.13< \Omega_{\rm m0}< 0.23$ from
supernova data and peaks in CMB power spectrum, $n\le 0.3$ from
WMAP data \cite{sen03}. Godlowski and Szydlowski found that
$\Omega_{\rm m0}=0.48^{+0.08}_{-0.13}$ and
$n=-0.4^{+0.77}_{-1.24}$, and $\Omega_{\rm
m0}=0.51^{+0.05}_{-0.06}$ and $n=-1.2^{+0.77}_{-1.06}$ from
different sets of supernova data \cite{smcw}. Frith found that
$0.19<\Omega_{\rm m0}<0.26$ and $0.01<n<0.24$ from supernova data
\cite{frith}. In general, different analysis gives different
results, but they are still consistent with each other at 99\%
confidence level. Dev, Alcaniz and Jain found that the best fit to
the gravitational lensing effect is $n=0.76$ and $\alpha=2.4$ by
assuming $\Omega_{\rm m0}=0.3$ for the generalized Cardassian
model \cite{dev03}. Taking the best fit result to WMAP data, we
get $z_{q=0}=0.82$.

For the DGP model, the best fit parameter to the WMAP TT and TE
data is $\Omega_{\rm m0}=0.174$ and the best fit parameter to the
ten supernova data is $\Omega_{\rm m0}=0.21$. Deffayet \etal found
that $\Omega_{\rm m0}=0.18^{+0.07}_{-0.06}$ from supernova
observations and $\Omega_{\rm m0}=0.3$ from CMB constraints
\cite{deffayet}. However, the constraint from supernova data and
CMB data are still consistent with each other at $1\sigma$ level.
If we take the bigger value $\Omega_{\rm m0}=0.21$, then we get
$z_{q=0}=0.81$. Multamaki, Gaztanaga and Manera considered the
effects of Cardassian model and DGP model on large scale structure
and found that these effects are different from that of
$\Lambda$-model \cite{maltamaki}.

Due to the uncertainties in the observational data, it is still
difficult to discriminate different dark energy models and
alternative models. Wang \etal showed that future supernova data
such as SNAP \footnote{http://snap.lbl.gov/} could differentiate
various models if $\Omega_{\rm m0}$ is known with 10\% accuracy
\cite{wang}. On the other hand, if we can determine the dynamical
evolution of equation of state parameter of dark energy with high
precision using model independent method, then we can discriminate
different models. Alam \etal showed that future SNAP data will
provide important insights on the nature of dark energy at high
redshifts \cite{alam}. Turner and Riess showed that $z_{q=0}\sim
0.5$ \cite{agr}, Daly and Djorgovski found that $z_{q=0} >0.3$
\cite{daly}. In conclusion, the above models are consistent with
current observations.

In terms of model building, we can assume a particular scale
factor which manifests early deceleration and later acceleration,
then find out the form of $g(x)$. For example, we take
$a(t)=a_0[\sinh(t/t_0)]^\beta/\alpha$, then we find
$g(x)=Ax^{2/3\beta}+B$. This model recovers the standard cosmology
when $\beta=2/3$.

\ack The author Gong is thankful to Padmanabhan T, Finelli F,
Godlowski W, Doran M and Sahni V for comments. The work of Gong is
fully financed by Chongqing University of Post and
Telecommunication under grants A2003-54 and A2004-05.

\end{document}